\documentclass[aip, amsmath,amssymb, reprint]{revtex4-1}
\usepackage{amsmath}
\usepackage{amssymb}
\usepackage{bm}
\usepackage{glossaries}
\usepackage{graphicx}
\usepackage{multirow}
\usepackage{fancyvrb}
\usepackage[linesnumbered,ruled]{algorithm2e}
\usepackage{siunitx}
\usepackage{subfigure}
\usepackage{threeparttable}

\usepackage[usenames,dvipsnames,svgnames]{xcolor}
\usepackage{hyperref}
\hypersetup{
pdfnewwindow=true,      % links in new window
colorlinks=true,        % false: boxed links; true: colored links
linkcolor=Blue,         % color of internal links
citecolor=Blue,         % color of links to bibliography
filecolor=Blue,         % color of file links
urlcolor=Blue           % color of external links
}

\newacronym{ace}{ACE}{atomic cluster expansion}
\newacronym{dft}{DFT}{density-functional theory}
\newacronym{eam}{EAM}{embedded-atom method}
\newacronym{dp}{DP}{deep potential}
\newacronym{gap}{GAP}{Gaussian approximation potential}
\newacronym{md}{MD}{molecular dynamics}
\newacronym{mc}{MC}{Monte Carlo}
\newacronym{mlp}{MLP}{machine-learned potential}
\newacronym{mtp}{MTP}{momentum tensor potential}
\newacronym{nep}{NEP}{neuroevolution potential}
\newacronym{tabgap}{tabGAP}{tabulated Gaussian approximation potential}
\newacronym{snes}{SNES}{separable natural evolution strategy}
\newacronym{rmse}{RMSE}{root-mean-square error}
\newacronym{zbl}{ZBL}{Ziegler-Biersack-Littmark}
\newacronym{hea}{HEA}{high-entropy alloy}
\newacronym{rhea}{RHEA}{refractory high-entropy alloy}
\newacronym{mpea}{MPEA}{multi-principal element alloy}
\newacronym{csro}{CSRO}{chemical short-range ordering}
\newacronym{sro}{SRO}{short-range order}
\newacronym{rss}{RSS}{random solid solution}
\newacronym{wc}{WC}{Warren-Cowley}
\newacronym{fp}{FP}{Frenkel pair}
\newacronym{nrt}{NRT-dpa}{Norgett-Robinson-Torrens displacements per atom}
\newacronym{arc}{arc-dpa}{athermal recombination corrected displacements per atom}
\newacronym{dpa}{dpa}{displacements per atom}
\newacronym{kp}{KP}{Kinchin-Pease}
\newacronym{sia}{SIA}{self-interstitial atom}
\newacronym{pka}{PKA}{primary knock-on atom}
\newacronym{bcc}{bcc}{body-centered cubic}
\newacronym{fcc}{fcc}{face-centered cubic}
\newacronym{msd}{MSD}{mean squared displacement}
\newacronym{w}{W}{tungsten}
\newacronym[longplural={threshold displacement energies}]{tde}{TDE}{threshold displacement energy}

\begin{document}

\title{Revealing the impact of chemical short-range order on radiation damage in MoNbTaVW high-entropy alloys using a machine-learning potential}

\author{Jiahui Liu}
%\email{b20200613@xs.ustb.edu.cn}
\affiliation{Beijing Advanced Innovation Center for Materials Genome Engineering, University of Science and Technology Beijing, Beijing 100083, China}
\affiliation{Institute for Advanced Materials and Technology, University of Science and Technology Beijing, Beijing 100083, China}

\author{Shuo Cao}
%\email{icaoshuo@outlook.com}
\affiliation{Beijing Advanced Innovation Center for Materials Genome Engineering, University of Science and Technology Beijing, Beijing 100083, China}
\affiliation{Institute for Advanced Materials and Technology, University of Science and Technology Beijing, Beijing 100083, China}

\author{Yanzhou Wang}
%\email{yanzhowang@gmail.com}
\affiliation{QTF Centre of Excellence, Department of Applied Physics, P.O. Box 15600, Aalto University, FI-00076 Aalto, Espoo, Finland}

\author{Zheyong Fan}
\email{brucenju@gmail.com}
 \affiliation{College of Physical Science and Technology, Bohai University, Jinzhou 121013, China}

\author{Guocai Lv}
%\email{lvguocai@ustb.edu.cn}
\affiliation{Basic Experimental Center of Natural Science, University of Science and Technology Beijing, Beijing 100083, China}

\author{Ping Qian}
%\email{qianping@ustb.edu.cn}
\affiliation{Beijing Advanced Innovation Center for Materials Genome Engineering, University of Science and Technology Beijing, Beijing 100083, China}

\author{Yanjing Su}
\email{yjsu@ustb.edu.cn}
\affiliation{Beijing Advanced Innovation Center for Materials Genome Engineering, University of Science and Technology Beijing, Beijing 100083, China}
\affiliation{Institute for Advanced Materials and Technology, University of Science and Technology Beijing, Beijing 100083, China}

\date{\today}

\begin{abstract}
The effect of chemical short-range order (CSRO) on primary radiation damage in MoNbTaVW high-entropy alloys is investigated using hybrid Monte Carlo/molecular dynamics simulations with a machine-learned potential. We show that CSRO enhances radiation tolerance by promoting interstitial diffusion while suppressing vacancy migration, thereby increasing defect recombination efficiency during recovery stage. However, CSRO is rapidly degraded under cumulative irradiation, with Warren-Cowley parameters dropping below 0.3 at a dose of only 0.03~dpa. This loss of ordering reduces the long-term enhancement of CSRO on radiation resistance. Our results highlight that while CSRO can effectively improve the radiation tolerance of MoNbTaVW, its stability under irradiation is critical to realizing and sustaining this benefit.

\end{abstract}

\maketitle

\Gls{w}-based \glspl{rhea} exhibit outstanding tolerance to high temperatures, stress, and irradiation, making them promising candidates for plasma-facing materials~\cite{Senkov2010intermetallics, Senkov2011intermetallics, Atwani2019sciadv, Atwani2023nc}.
This class of \glspl{hea}, composed of multiple principal elements at high concentrations, provides unprecedented flexibility in material design~\cite{Yeh2004, Miracle2017acta}. 
The resulting compositional complexity often promotes \gls{csro}~\cite{Bo2021jacs, Walsh2023mrs, Xu2023apl, Sheriff2024pnas}, which can profoundly influence their mechanical and functional properties~\cite{George2019nrm, Pedro2023acta,Jian2024apl, Zhao2024acta, Zhao2024npjmd}.

Given the experimental challenges in probing \gls{csro}, \gls{md} and \gls{mc} simulations have become essential tools for investigating its atomic-level impact on radiation damage.
Liu \textit{et al.} employed hybrid MC/MD simulations to examine the role of \gls{csro} in radiation resistance~\cite{Liu2023jmst, Liu2024pns}. 
Their results reveal that \gls{csro} increases the migration energies of interstitials and vacancies, thereby suppressing defect migration and aggregation. 
Li \textit{et al.} studied primary radiation damage in the Al$_{0.5}$CoCrFeNi alloy and found that \gls{csro} effectively suppresses point defect generation by increasing the formation energies of vacancies and interstitials, as well as the migration barriers of vacancies~\cite{Li2024mtc}. 
Additionally, Arkoub \textit{et al.} used \gls{md} simulations to demonstrate that \gls{csro} in Fe-Ni-Cr alloys evolves dynamically upon irradiation, either decreasing or increasing before reaching a steady-state value~\cite{Arkoub2023scripta}.
To date, most perspectives regarding the influence of \gls{csro} on radiation damage have focused on \gls{fcc} \glspl{mpea}.
There is still a lack of large-scale atomistic simulations of radiation damage in \glspl{rhea}, with existing studies~\cite{Li2020npj,Byggmastar2021prb, jesper2024prm} primarily addressing the role of \gls{csro} in segregation and \gls{tde}.
This is partly due to the need for larger systems and longer simulations in hybrid MC/MD studies of \glspl{rhea}, which substantially increases the computational cost.

Recently, a canonical-ensemble MC/MD algorithm has been implemented in the open-source \textsc{gpumd}~\cite{Fan2017cpc} package. 
In this approach~\cite{Song2024arx}, the computational cost of each MC trial is independent of the system size and depends only on the average number of atomic neighbors. 
For sufficiently large systems (e.g., 490,000 atoms with an average coordination number $M \approx 60$), the computational cost of MC steps becomes negligible, allowing the overall simulation speed to approach that of pure MD. 
This paves the way for large-scale simulations of the impact of \gls{csro} on primary radiation damage.

In this study, all MC/MD simulations were performed using \textsc{gpumd} with a \gls{nep} model \cite{fan2021prb, song2024nc} constructed for Mo-Nb-Ta-V-W systems~\cite{Liu2024arx} based on the training dataset of a \gls{tabgap} model~\cite{Byggmastar2020prm, Byggmastar2021prb, Byggmastar2022prm}.
The \gls{nep} approach is a highly computationally efficient approach that has found widespread applications in compositionally complex materials \cite{ying2025cpr}.
To validate its accuracy, we first reproduced \gls{csro}-related simulations previously conducted with the \gls{tabgap} model, including (i) ordering and segregation in a single-crystal equiatomic \gls{rhea}, and (ii) threshold displacement energy calculations.
As shown in Supplementary Fig.~1, Mo and Ta exhibit local ordering at low temperatures. 
Mo–Nb and W–V first-nearest-neighbor (1NN) pairs are also favored, with W–V additionally showing second-nearest neighbor (2NN) attraction, suggesting more complex local structures. 
Above 300~K, the ordered phases gradually transition into a \gls{rss}.
Supplementary Fig.~2 shows the angular maps of \glspl{tde} for the five \gls{pka} types in the SRO-HEA, with values of 60 eV (V), 62 eV (Nb), 55 eV (Mo), 52 eV (Ta), and 50 eV (W), averaging 56 eV. 
Supplementary Fig.~3 compares the distributions between the SRO-HEA and the reference RSS-HEA~\cite{Liu2024arx}. 
\gls{csro} leads to an upward shift of about 5 eV in both the distribution and average.
All results above are consistent with previous findings based on the \gls{tabgap} model~\cite{Byggmastar2021prb, jesper2024prm}.
Additional computational details are provided in the Supplementary Notes.

\begin{figure}[htp] 
\centering 
\includegraphics[width=\columnwidth]{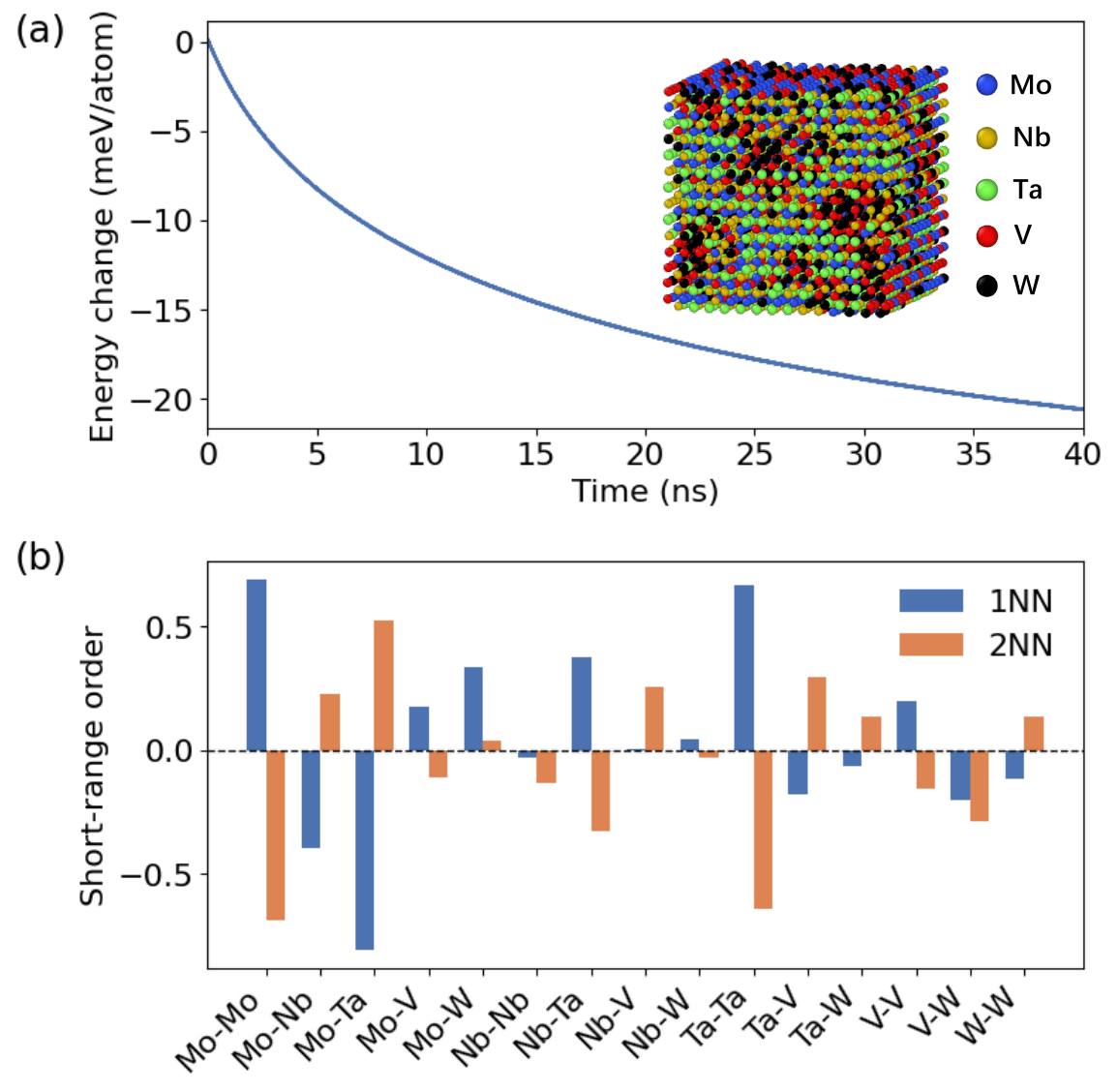}
\caption{(a) Potential energy change from the random alloy to an ordered structure as a function of MD simulation time. (b) First-nearest-neighbor and second-nearest-neighbor Warren–Cowley parameters in the short-range ordered alloy. The inset in (a) shows a representative local atomic configuration exhibiting short-range order.} 
\label{fig:sro}
\end{figure}

After evaluating the reliability of the \gls{nep} model in describing \gls{csro} and atomic displacements, we prepared a large SRO-HEA system with simulation box dimensions of $150a_0 \times 150a_0 \times 150a_0$ (6,750,000 atoms) to investigate the impact of \gls{csro} on radiation damage in MoNbTaVW \glspl{hea}.
Figure~\ref{fig:sro}(a) shows the potential energy as a function of time during hybrid MC/MD equilibration under the isothermal-isobaric ensemble at 300 K, with 100 MC trails every 10 MD steps. 
Figure~\ref{fig:sro}(b) presents the 1NN and 2NN \gls{wc} parameters in the final configuration. 
The \gls{wc} parameter~\cite{WC_parameters} for an element pair AB is computed as
\begin{equation}
\alpha_{AB} = 1 - \frac{p^{AB}}{c_B}, 
\end{equation}
where $p^{AB}$ is the probability of finding a B atom at a nearest-neighbor site of an A atom, and $c_B$ is the concentration of element B in the alloy. 
A negative $\alpha_{AB}$ suggests a tendency toward ordering between A and B atoms, while a positive value means repulsion.
Although the larger system size here leads to a slight reduction in \gls{wc} parameters compared to smaller system size, the chemical ordering is maintained, confirming that the 40~ns simulation is sufficient to establish well-developed \gls{csro}.
The inset in Fig.~\ref{fig:sro}(a) shows a local atomic snapshot, in which Mo-Ta pairs exhibit an almost perfect B2-type ordering.

\begin{figure}[h]
\centering
\includegraphics[width=0.95\columnwidth]{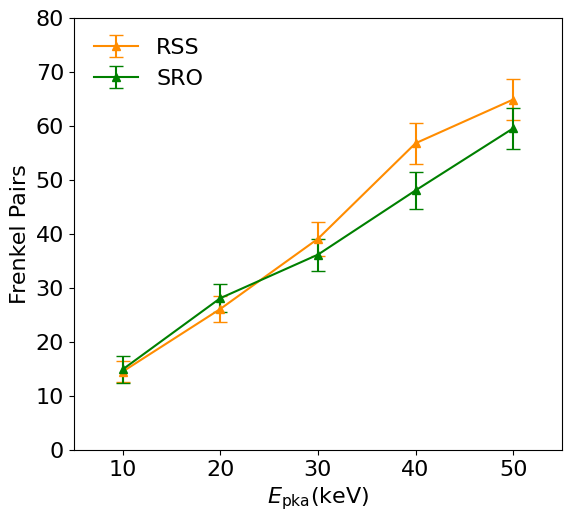}
\caption{The residual point defects in random and short-range ordered alloys. Error bars represent the standard deviations from 40 independent simulations.}
\label{fig:fps_final}
\end{figure}

Displacement cascade simulations were then performed on both \gls{csro} and \gls{rss} MoNbTaVW \glspl{hea} at \gls{pka} energies of 10, 20, 30, 40, and 50~keV. 
To ensure statistical convergence and stable results, 40 simulations were carried out for each \gls{pka} energy, each lasting about 50~ps.
For the SRO-HEAs, the prepared configuration was randomly shifted before each recoil event to sample different local chemical environments.
For the RSS-HEAs, a same-sized random equimolar \gls{bcc} cell was generated and equilibrated at 300~K and 0~GPa for 30~ps prior to each simulation.
The central atom was selected as the \gls{pka} and given an initial velocity in a random direction. 
A Nosé–Hoover chain thermostat~\cite{Martyna1992jcp} was applied to a 10~\AA{}-thick boundary region, while all other atoms evolved under the NVE ensemble.
The time step was dynamically adjusted to limit the maximum atomic displacement to 0.015~\AA\ per step, with an upper bound of 1~fs.
Electronic stopping \cite{Kai1995cms} was applied as a frictional force on atoms with a kinetic energy over 10 eV, using data from the SRIM-2013 code \cite{ziegler2010srim, ziegler2013}. 
All reported results are statistical averages over the 40 independent simulations.

\begin{figure}[h]
\centering
\includegraphics[width=\columnwidth]{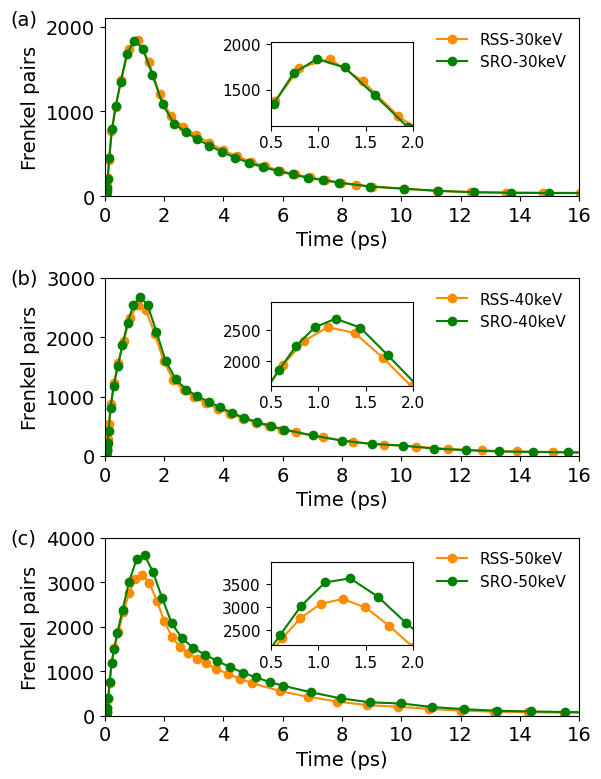}
\caption{The number of Frenkel pairs as a function of simulation time in random and short-range ordered alloys during cascade simulations at (a) 30 keV, (b) 40 keV, and (c) 50 keV.}
\label{fig:fps_evo}
\end{figure}

Figure~\ref{fig:fps_final} shows the average number of \glspl{fp} that survived at the final stage of the cascade simulations in both the SRO-HEAs and RSS-HEAs. 
Although \gls{csro} increases the \gls{tde}, the number of \glspl{fp} at lower energies (10 and 20~keV) remains comparable between the two systems. 
As the \gls{pka} energy increases, SRO-HEAs exhibit a notable suppression in the number of surviving point defects.
Figure~\ref{fig:fps_evo} presents the time evolution of the number of \glspl{fp} during the cascade process at \gls{pka} values of 30, 40, and 50~keV. 
At higher energies, more \glspl{fp} are generated in SRO-HEAs during the thermal spike stage, possibly due to less efficient energy dissipation in chemically ordered regions, which leads to a larger molten zone in the cascade core.
The energy loss to electronic interactions, listed in the Supplementary Table~1, is nearly identical between the SRO-HEA and RSS-HEA systems across all \gls{pka} energies.
Taken together, the increased defect recombination and comparable electronic energy loss indicate that the thermal relaxation phase plays a significant role.
Moreover, Supplementary Fig.~4 shows the \gls{wc} parameters in the molten region after cooling. 
The persistence of \gls{csro} likely affects the migration of interstitial atoms and the recombination of \glspl{fp} during the recovery stage.

\begin{figure*}[t]
\centering
\includegraphics[width=1.6\columnwidth]{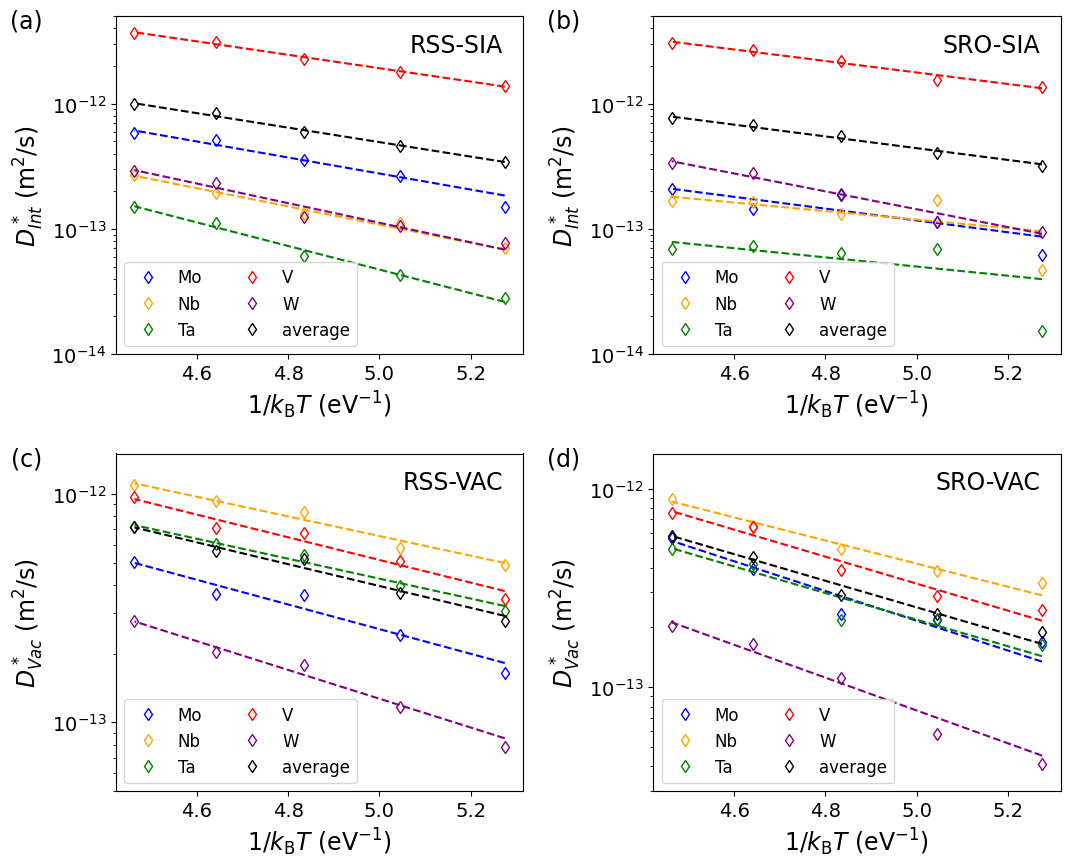}
\caption{Tracer diffusion coefficients of elements for interstitial diffusion in (a) random and (b) short-range ordered alloys, and for vacancy diffusion in (c) random and (d) short-range ordered alloys.}
\label{fig:diffusion}
\end{figure*}

\begin{figure}[h]
\centering
\includegraphics[width=\columnwidth]{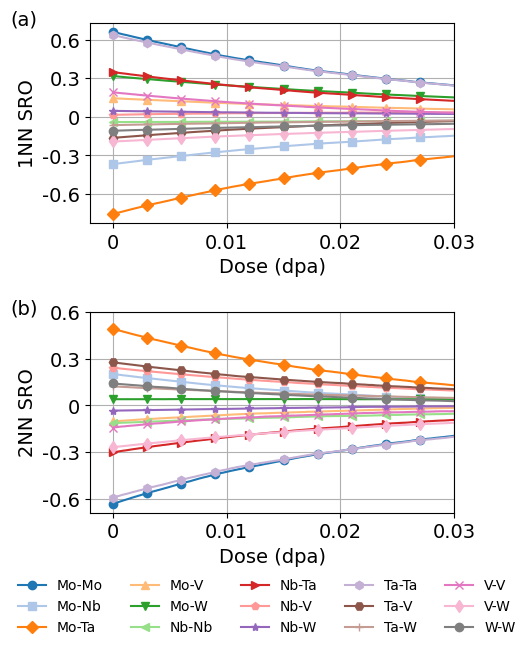}
\caption{Evolution of (a) first-nearest-neighbor and (b) second-nearest-neighbor Warren–Cowley parameters with dose in the short-range ordered alloy.}
\label{fig:dose}
\end{figure}

To further investigate defect evolution during the recovery stage, we simulated the diffusion of a single interstitial and a single vacancy in both SRO-HEAs and RSS-HEAs. 
Equiatomic random solid solutions of $16a_0 \times 16a_0 \times 16a_0$ unit cells (8192 atoms) were constructed, with the SRO system equilibrated using hybrid MC/MD at 300~K for 1~ns.
An interstitial was introduced by adding one extra W atom to the perfect cell (8193 atoms), and a vacancy was created by removing one random atom (8191 atoms). 
Prior to diffusion simulations, the SRO and RSS systems were equilibrated at temperatures ranging from 2200~K to 2600~K (in 100~K intervals) and zero pressure for 10~ps in the NPT ensemble.
Diffusion was simulated for 10~ns in the NVE ensemble with a correlation time of 1~ns, and both average and elemental \glspl{msd} were computed.
The temporal evolution of elemental and average \glspl{msd} for interstitial and vacancy diffusion is presented in Supplementary Fig.~5 and 6.
For interstitial diffusion, V atoms dominate the overall migration, especially in the SRO-HEAs, where the diffusion of Nb, Ta, and Mo atoms becomes markedly non-linear. 
It is worth noting that V atoms are the primary interstitial defects generated during the radiation damage of MoNbTaVW.
In the case of vacancy diffusion, all \glspl{msd} increase approximately linearly with time. 
However, in SRO-HEAs, vacancy migration is noticeably suppressed.

Subsequently, the tracer diffusion coefficients $D^*$ were calculated from the slope of the \gls{msd} as a function of time, according to the Einstein relation~\cite{Einstein1905adp}:
\begin{equation}
D^* = \frac{\mathrm{MSD}(t)}{6t}.
\end{equation}
Figure~\ref{fig:diffusion}(a) and (b) show the average and elemental tracer diffusion coefficients for interstitials in the RSS-HEAs and SRO-HEAs.
In the RSS-HEAs, the elemental tracer diffusion coefficients follow the order $\mathrm{V} > \mathrm{Mo} > \mathrm{W} \approx \mathrm{Nb} > \mathrm{Ta}$, 
and only the interstitial tracer diffusion coefficient ($D^*_{\mathrm{Int}}$) of V exceeds the average value across the temperature range.
Notably, in the SRO-HEAs, the diffusion of Mo, Nb, and Ta is significantly suppressed, likely due to the formation of Mo-Ta and Mo-Nb \gls{csro}.
Figure~\ref{fig:diffusion}(c) and (d) show the average and elemental tracer diffusion coefficients for vacancy diffusion in the RSS-HEAs and SRO-HEAs.
The vacancy tracer diffusion coefficient ($D^*_{\mathrm{Vac}}$) values of V and Nb are higher than the average ones in both systems. 
In contrast, the tracer diffusion coefficients of Mo and Ta in the SRO-HEAs exhibit considerable fluctuations, as \gls{csro} affects vacancy diffusion.

The migration energy $E_m$ was obtained by fitting the Arrhenius equation~\cite{Zhang2022acta, Kulitckii2023acta}:
\begin{equation}
D^* = D_0 e^{-\frac{E_m}{k_\text{B} T}},
\end{equation}
where $D_0$ is the pre-exponential factor, $k_\text{B}$ is the Boltzmann constant and $T$ is the temperature.
The migration energies of interstitials and vacancies in the RSS-HEA are $E_m^\mathrm{int} = 1.33 \pm 0.08$~eV and $E_m^\mathrm{vac} = 1.10 \pm 0.11$~eV, which are consistent with the values reported using the \gls{tabgap} model~\cite{Wei2024acta}.
In contrast, the migration energies in the SRO-HEA are $E_m^\mathrm{int} = 1.08 \pm 0.08$~eV and $E_m^\mathrm{vac} = 1.54 \pm 0.14$~eV.
The presence of \gls{csro}, particularly the formation of Mo-Ta and Mo-Nb pairs, effectively influences defect diffusion by enhancing interstitial migration while suppressing vacancy migration.
This imbalance leads to more frequent recombination of point defects during the recovery stage, thereby improving the material's resistance to irradiation damage.

Although the \gls{csro} is maintained during the primary radiation damage, the inherent tendency of MoNbTaVW to transform toward a \gls{rss} configuration at elevated temperatures may lead to its degradation as the radiation dose accumulates.
To evaluate the stability of the \gls{csro}, a series of overlapping cascades were simulated as cumulative single-cascade events, each followed by a brief relaxation run.
The dose after $N$ cascades, in units of \gls{dpa}, was calculated according to the \gls{nrt} model~\cite{norgett_proposed_1975}, using a damage energy of $T_d = 10$~keV. 
The simulation box, with dimensions of $100a_0 \times 100a_0 \times 100a_0$ (2,000,000 atoms), was sufficiently large to prevent cascades from interacting with the simulation boundaries in SRO-HEAs.
A total of 1000 cascades were simulated sequentially, resulting in an accumulated dose of approximately 0.03~\gls{dpa}, which corresponds to a relatively low irradiation dose.
Figure~\ref{fig:dose} shows the \gls{wc} parameters as a function of dose for both (a) 1NN and (b) 2NN pairs.
As shown, the \gls{csro} is rapidly degraded with increasing dose. At a dose of 0.03~\gls{dpa}, the absolute values of the \gls{wc} parameters drop below 0.3, indicating a weak level of chemical ordering.
This highlights that although \gls{csro} enhances radiation resistance, maintaining its stability under increasing radiation dose is essential to sustaining this improvement.

In summary, our hybrid MC/MD simulations with the neuroevolution potential provide novel insights into how \gls{csro} influences radiation damage behavior in MoNbTaVW \glspl{hea}.
The presence of \gls{csro} enhances radiation resistance by facilitating interstitial diffusion and suppressing vacancy migration at higher \gls{pka} energies, thereby increasing the defect recombination efficiency.
However, the \gls{csro} is rapidly degraded with increasing radiation dose. 
Thus, utilizing \gls{csro} to enhance the radiation tolerance of MoNbTaVW is a viable strategy, provided that its stability can be improved. This may be achieved, for instance, through the introduction of grain boundaries or phase interfaces.

\textbf{Data availability}

The data presented in this paper are available from the corresponding authors upon reasonable request.

\textbf{Acknowledgments}
This work was supported by the National Science and Technology Advanced Materials Major Program of China (No. 2024ZD0606900), and USTB MatCom of Beijing Advanced Innovation Center for Materials Genome Engineering. 

\bibliography{myrefs.bib}

\end{document}